%% file: main.tex
\documentclass[sigconf,authorversion=true]{acmart}

\AtBeginDocument{%
  \providecommand\BibTeX{{%
    \normalfont B\kern-0.5em{\scshape i\kern-0.25em b}\kern-0.8em\TeX}}}

\setcopyright{acmcopyright}
\copyrightyear{2023}
\acmYear{2023}
\acmDOI{XXXXXXX.XXXXXXX}
\acmDOI{}

\acmConference[ISPD 2023]{International Symposium on Physical Design}{March 26--29, 2023}{Online}
\acmPrice{15.00}
\acmISBN{978-1-4503-XXXX-X/18/06}

\usepackage{soul}
\usepackage{multirow}
\usepackage{ragged2e}
\usepackage[noend]{algpseudocode}     
\usepackage{algorithm}            
\usepackage{threeparttable}
\usepackage{subfig}
\newcommand{\tsc}[1]{\textsuperscript{#1}} 

\algrenewcommand\algorithmicforall{\textbf{foreach}}

\newcommand{\smallerspace}{\vspace{-0.5em}}
\newcommand{\littlesmallerspace}{\vspace{-0.25em}}

\newcommand{\TM}{\textit{TroMUX}}

\newcommand{\belowfigurespace}{\vspace{-10pt}}
\newcommand{\belowtablespace}{\vspace{-5pt}}
\renewcommand{\belowfigurespace}{\vspace{-0pt}}
\renewcommand{\belowtablespace}{\vspace{-0pt}}

\newcommand{\ull}[1]{\ul{#1}}

\begin{document}

\author{Fangzhou Wang\tsc{1}, Qijing Wang\tsc{1}, Bangqi Fu\tsc{1}, Shui Jiang\tsc{1}, Xiaopeng Zhang\tsc{1}, Lilas Alrahis\tsc{2}, \\Ozgur Sinanoglu\tsc{2}, Johann Knechtel\tsc{2}, Tsung-Yi Ho\tsc{1}, and Evangeline F. Y. Young\tsc{1}}
\email{{fzwang,qjwang21,bqfu21,sjiang22,xpzhang}@cse.cuhk.edu.hk} 
\email{{lma387,ozgursin,johann}@nyu.edu}
\email{{tyho,fyyoung}@cse.cuhk.edu.hk}
\affiliation{%
  \country{\tsc{1}The Chinese University of Hong Kong, \tsc{2}New York University Abu Dhabi}
}

\renewcommand{\shortauthors}{Wang et al.}

%



\title{Security Closure of IC Layouts Against Hardware Trojans}

\begin{abstract}
\input{abstract}
\end{abstract}

\maketitle

\section{Introduction}
\label{sec:intro}

In this work, we focus on the threat of {hardware Trojans}, i.e., malicious circuit modifications.
By design, Trojans are only minor in extent but severe in
fallout~\cite{xiao16,dong20,yang16_a2}.
Many Trojan countermeasures have been proposed over the years (see also Sec.~\ref{sec:bg:prior} and \ref{sec:prior}).
Aside from
(i)~reactive, post-silicon monitoring
and (ii)~pre-/post-silicon verification, testing, detection, and inspection,
we argue that (iii)~proactive, pre-silicon prevention
is essential to hinder Trojans to begin with.
However, most related prior art falls short in terms of resilience against advanced attacks and/or overheads.

The \textbf{objective} of this work is security closure against hardware Trojans.
Note that security closure is an emerging paradigm to proactively harden the physical layouts of integrated circuits (ICs) at
design-time against various threats that are executed post-design time~\cite{knechtel21_SCPL_ICCAD, knechtel22_SCPL_ISPD}.
In this work, we
aim for \textit{proactive, pre-silicon Trojan prevention, by carefully and systematically hardening the physical
layout of ICs as a whole against post-design Trojan insertion}.

Toward that end, and more so for Trojan defense in general, we have identified the following
\textbf{research challenges}:
\begin{itemize}

\item \textit{Robustness -- Any defense
must remain robust in place.}
That is, a foundry-based adversary, being fully aware that some Trojan defense is put in
place, would naturally want to first circumvent that defense
(i.e., stealthily remove, override, or otherwise render useless)
before inserting their Trojan.
\ull{Such second-order attacks represent a key challenge where
prior art, be they proactive or reactive defenses, falls short.}

\item \textit{Effectiveness -- Any defense should be able to protect against various Trojans.}
This is especially true for proactive, pre-silicon schemes which have
only ``one shot'' at design-time.
     \ull{Layouts should be protected as a whole}, i.e., in terms of (i)~layout resources
that are exploited for Trojan insertion (e.g., open placement sites, free routing tracks, available timing
	slacks) and (ii)~structure and functionality,
analysed by adversaries for targeted Trojan insertion.

\item \textit{Efficiency -- Any defense should incur limited, controllable overheads.}
Taking control over such trade-offs requires two parts:
\ull{(i)~metrics for security and overheads, and (ii)~some integrated, secure-by-design methodology,}
utilizing the metrics and user guidance as needed.

\end{itemize}

\noindent Accordingly, our work makes the following
\textbf{contributions}:
\begin{enumerate}

\item Layout-level logic locking
--
We propose a multiplexer (MUX)-based locking scheme, called \TM, devised to hinder post-design Trojan
insertion.
Our locking scheme addresses the above challenges through the following capabilities:
	\begin{enumerate}

	\item Robustness --
	\ull{\textit{TroMUX} is devised to withstand state-of-the-art, machine learning (ML)-based attacks on
	locking}, like \textit{SCOPE}~{\cite{alaql21}} and \textit{MuxLink}~{\cite{alrahis22_ML}}.
	This is essential to hinder adversaries from circumventing the Trojan defense.

	\item Effectiveness --
	By the security promise of locking, \ull{attackers
	cannot easily insert targeted Trojans anymore}, i.e., Trojans that require understanding of the original design.
	Furthermore,
	by densely filling up the
	layout with as many \TM\ instances as practically possible (keeping the design quality well under control),
	\ull{attackers cannot easily insert additional Trojan logic in general anymore.}

	\end{enumerate}

\item Integrated security closure --
	We propose an effective and efficient methodology for security closure against post-design Trojan insertion. The methodology is fully integrated
	into a commercial-grade physical-synthesis flow.
	Such integration is essential to achieve (i) the above outlined security principles and (ii) take control of
	security-versus-overheads trade-offs arising for security closure of layouts.

\end{enumerate}

\textbf{Release:}
We will release 
our source codes.
We release our artifacts of secured layouts already during pre-publication~\cite{release}, to
enable independent verification of our work.

\section{Background and Motivation}
\label{sec:bg}

\subsection{Trojans}
\label{sec:bg:Trojans}

\subsubsection{Working Principles}

Trojans are malicious hardware modifications~\cite{xiao16,dong20}.
This notion is very diverse, covering modifications that:
(i)~leak information from an IC, reduce its performance, or disrupt its working altogether;
(ii)~are always on, triggered internally, or triggered externally;
(iii)~are introduced through untrustworthy third-party intellectual property (IP), adversarial designers, during outsourced mask generation,
manufacturing, packaging of ICs; etc.

Most if not all Trojans comprise a trigger and a payload; the trigger
activates the payload on some attack condition, and the payload serves to perform an actual attack.
Triggers are often based on low-controllability nets (LCNs)---to complicate their detection during testing---whereas
payloads are targeting on sensitive assets like key registers.
Note that trigger and payload are implemented individually but are working together in tandem.
Also note that most Trojans require some layout-level resources like open placement sites, free routing
tracks, and/or available timing slacks.

\subsubsection{Prior Art for Countermeasures}
\label{sec:bg:prior}

Prior art for Trojan countermeasures can be classified into:
\begin{itemize}
\item Proactive detection schemes, that is
	pre-silicon verification
	(e.g., \cite{dong20,chen18}),
	post-silicon testing (e.g., \cite{dong20,deng20}), or post-silicon inspection (e.g.,
			\cite{vashistha18});
\item Proactive, pre-silicon prevention schemes
	(e.g.,
	 \cite{dupuis14,marcelli17,samimi16,sisejkovic19,xiao14,ba15,ba16,knechtel21_SCPL_ICCAD,hosseintalaee17});
\item Reactive, post-silicon monitoring schemes
	(e.g., \cite{guimaraes17,hou18,vijayan20}).
\end{itemize}
Proactive detection and reactive monitoring schemes are
generally challenged by advanced and stealthy Trojans~\cite{yang16_a2, guo19}.
Further, reactive monitoring as well as proactive prevention schemes typically require some dedicated hardware support---if not secured properly,
{the related circuitry may well be circumvented by adversaries during the course of Trojan insertion}.

Given these challenges, {a \textit{robust} scheme for proactive prevention is
essential to hinder Trojans early on, as best as possible.}
We discuss prior art for proactive, pre-silicon prevention in more detail in Sec.~\ref{sec:prior}.
A comparison of prior art and ours is also outlined in Table~\ref{tab:hl_comp}.

Finally, note that the three classes are orthogonal, yet generally compatible.
To increase the overall resilience against Trojans, techniques from all classes could (and should) be utilized
jointly, if permissible in terms of design overheads and cost.

\subsection{Security Closure}
\label{sec:bg:SCPL}

As indicated,
security closure is an emerging paradigm that
seeks to proactively harden the physical layouts of ICs,
at design time, against various threats that are executed post-design time~\cite{knechtel21_SCPL_ICCAD,
knechtel22_SCPL_ISPD}.
In the broader context of secure-by-design efforts for electronic design automation (EDA) tools~\cite{
	sigl11, ravi19, wu21},
security closure aims for secure physical-synthesis stages.

Security closure against Trojans means to control physical synthesis such that insertion of Trojans
becomes impractical, while at the same time managing the impact on design quality of such
measures~\cite{knechtel21_SCPL_ICCAD,knechtel22_SCPL_ISPD}.
For example, an aggressively dense layout would leave only few open placement sites
and few routing resources
exploitable for Trojan insertion.
{While aggressively dense layouts are already challenging by themselves, in terms of managing design quality, such naive approach is still not
good enough for security closure.}
This is because, for one, an advanced Trojan like \textit{A2}~\cite{yang16_a2} may require only 20 placement sites for
an advanced analog
implementation~\cite{trippel20};\footnote{This is a remarkable exception---other Trojans reported in
	the literature, as well as a digital version of \textit{A2} itself, require hundreds or thousands of
		sites~\cite{trippel20}.}
such very few sites are likely to remain even in aggressively dense layouts.
For another, imagine a second-order attack where an adversary
would first revise the layout as much as necessary
(but also as little as possible, to avoid subsequent detection),
and only then insert their Trojan.

In short, {successful efforts for security closure against Trojans need to holistically address the challenges outlined in
	{Sec.~\ref{sec:intro}}}.

\input{incl/tables/table_prior}

\subsection{Logic Locking}
\label{sec:bg:LL}

Logic locking, or locking for short, means to
incorporate so-called key-gate structures
that are controlled by secret key-bits.
While locking is largely known for protection of IC's
IP, it can also
serve for Trojan defense.
In fact, different such schemes have been proposed: AND/OR locking~\cite{dupuis14}, X(N)OR
locking~\cite{samimi16,marcelli17}, and custom locking~\cite{sisejkovic19}.
	We discuss relevant prior art in more detail in Sec.~\ref{sec:prior:locking}.

Recently, {ML-based attacks
	like \textit{SAIL}~{\cite{chakraborty18}}, \textit{SCOPE}~{{\cite{alaql21}},
	\textit{OMLA}~{\cite{alrahis21_OMLA}} and \textit{MuxLink}~\cite{alrahis22_ML}
succeeded in learning various design features and subsequently predicting key-gates, all in an oracle-less setting,\footnote{%
	Only the oracle-less setting is relevant for our work as we are utilizing locking for proactive, pre-silicon
			prevention of Trojans, not for IP protection against end-users.}
	  i.e., without need for a functional chip that can be queried for its functional behaviour.}
For example, \textit{MuxLink}
learns on the structure of regular, unlocked parts of the design at hand, using a graph representation of the design,
and then deciphers the
MUX key-gates.
More specifically, \textit{MuxLink} considers each MUX key-gate's input at a
time, connecting it to the output of the MUX,
thereby essentially bypassing the key-gate, and contrasts the
resulting different structures to the trained knowledge to predict the
more likely structure.

In short, {ML-resilient locking schemes remain an open challenge, but are essential
for locking-based, proactive Trojan prevention.}
 
\section{Prior Art}
\label{sec:prior}

\subsection{Locking-Based Trojan Prevention}
\label{sec:prior:locking}

Dupuis et al.~\cite{dupuis14} lock LCNs using AND/OR key-gates, to hinder insertion of
Trojan triggers.
They consider timing slacks, varying toggling thresholds, and
balanced switching probabilities.
Samimi et al.~\cite{samimi16} follow similar principles as Dupuis et al., but utilize X(N)OR
key-gates.
Marcelli et al.~\cite{marcelli17} propose a multi-objective
algorithm,
seeking to
minimize LCNs and maximize the efficacy of X(N)OR locking at the same time.
\v{S}i\v{s}ejkovi\'{c} et al.~\cite{sisejkovic19} secure inter-module control signals against
software-controlled hardware Trojans, using encryption circuitry along with regular locking.
For the latter, they do not specify/limit the type of key-gates.

The above prior art has limitations as follows.
\begin{itemize}
\item Dupuis et al.~\cite{dupuis14} cannot protect against insertion of targeted Trojans. This is because they
utilize AND/OR key-gates to tune the controllability of nets, not to obfuscate the design.
\item \v{S}i\v{s}ejkovi\'{c} et al.~\cite{sisejkovic19} only protect against Trojans targeting on
inter-module control signals, not on modules' internals.
\item None consider Trojan prevention at the layout level; all are working only at netlist
level. This implies that none can conclusively prevent insertion of Trojan logic into the layout.
\item None explicitly prevent Trojan payloads.
\v{S}i\v{s}ejkovi\'{c} et al.~\cite{sisejkovic19}
are using locking in general, without focus on triggers or payloads, and others are locking
only low-controllability nets (to hinder insertion of Trojan triggers).
\item None show robustness against ML-based attacks;
	all are prone to second-order attacks bypassing the locking defense.
This is most evident for Dupuis et al.~\cite{dupuis14}: they employ a direct, hard-coded
correlation of key-bit `0' to OR key-gates, key-bit `1' to AND key-gates, respectively, and do not employ
re-synthesis, which is essential for obfuscation of that correlation, but would be challenging for their objective of
tuning the controllability of nets.
\end{itemize}
	{Note that, in contrast, our work addresses all these limitations.}

\subsection{Layout-Level Trojan Prevention}
\label{sec:prior:layout}

Xiao et al.~\cite{xiao14} fill layouts with built-in self-test components, arguing that tampering of
those structures (by adversaries regaining layout resources to insert their Trojans) is detected by
post-silicon testing.
Similarly, Ba et al.~\cite{ba15,ba16} fill layouts with additional circuitry.
Knechtel et al.~\cite{knechtel21_SCPL_ICCAD} propose techniques for Trojan-resistant physical synthesis,
based on locally increasing placement density (to hinder Trojan insertion) and locally increasing routing
density (to hinder Trojan routing).
Hossein-Talaee et al.~\cite{hosseintalaee17} redistribute white space/open sites otherwise exploitable for
Trojan insertion.

The above prior art has limitations as follows.
\begin{itemize}
\item None protects against insertion of targeted Trojans. In the absence of
locking or other obfuscation schemes, the original design is fully accessible by adversaries.
\item None conclusively shows robustness against second-order attacks,
as in adversaries regaining layout resources
despite the defense put in place.
For example, for the work in \cite{hosseintalaee17},
shifting of white spaces can be trivially reverted.
\item For full efficacy,
	the work in \cite{xiao14} requires 100\% test coverage
which can be difficult to achieve.
Further, the methodology is challenged by high utilization rates, limiting its practicability.
\item For the work in \cite{xiao14,ba15,ba16}, the number of additionally required primary inputs
scales with layout filling.
This is impractical; pads for primary inputs/outputs (PIs/POs) are large in actual ICs and, if not employed wisely, can
considerably increase the chip outline,
directly increasing cost for silicon area.
\end{itemize}
	{Note that, in contrast, our work addresses all these limitations.}

\section{Threat Models}
\label{sec:tm}

\textbf{Trojans:}
We follow a classical threat model that is consistent with the literature as follows.
\begin{enumerate}

\item We assume adversaries reside within manufacturing facilities, whereas the design process is considered 
trustworthy.

\item From (1) follows that adversaries have no knowledge of the original unprotected design, only of the
protected physical layout at hand.

\item  From (1) also follows that we do not account for Trojans introduced by, e.g., malicious designers or third-party IP
modules. Against such threats, and to increase a layout's overall Trojan resilience, countermeasures orthogonal to
ours, namely proactive detection or reactive monitoring schemes
(Sec.~\ref{sec:bg:prior}), may be applied along with ours.

\item We assume Trojans are implemented using regular standard cells, not by modifying interconnects or
transistors, etc.

\item We assume targeted Trojans with triggers based on LCNs and payloads targeting
on assets, i.e., security-critical components like key registers.

\item The adversaries' objective is to insert
some Trojan(s), more specifically trigger and payload components,
into the layout.

\end{enumerate}

\textbf{Locking:}
First, recall that we employ locking not for IP protection, but to hinder Trojan insertion.  Toward that end, we
follow an oracle-less threat model that is consistent with both the literature and the above threat model on
Trojans as follows.
\begin{enumerate}

\item The adversaries' objective is to circumvent the locking scheme, to (a) understand the true functionality of the
design, required for targeted Trojan insertion, and (b) to regain layout resources, required for Trojan insertion in general.

\item Given that adversaries are assumed to reside within manufacturing facilities, only oracle-less attacks
are applicable. None of the well-known Boolean satisfiability-based attacks are applicable.
Neither applicable are oracle-less attacks on locking schemes unlike \TM, e.g., \cite{yasin17_Anti-SAT}.

\item Following Kerckhoffs' principle, all implementation details for \TM\ locking are known to the adversaries;
only the key-bits remain undisclosed.

\end{enumerate}

\section{Methodology}
\label{sec:method}

This work is motivated by the need for a proactive, pre-silicon Trojan-prevention scheme that is robust, effective,
and efficient. As discussed in Sec.~\ref{sec:prior}, prior art falls short toward that end.

Our approach---which can be summarized as locking and layout filling applied in unison---aims to hinder first-order
and second-order Trojan-insertion attacks.
We integrate an ML-resilient, MUX-based locking scheme
directly into the IC layouts. Unlike prior art, we neither
employ trivial filler/spare cells, nor separate circuitry, nor vulnerable locking schemes.
To protect IC layouts holistically, our methodology carefully embeds, in a security-aware manner, as many locking
instances during physical instances as practically possible, i.e., keeping the design quality well under control.
Our methodology is fully integrated into commercial-grade design tools.

Next, we describe the components of our methodology, i.e., a MUX-based locking scheme and a
physical-synthesis flow for security- and design-aware locking and layout filling.
Note that both are devised to be working in unison, but also require individual
solutions toward that end.

\subsection{\TM}

Our MUX-based locking scheme, \TM, is specifically devised to hinder post-design Trojan
insertion.
Given the rise of powerful ML-based attacks on simple locking schemes, e.g., on X(N)OR
key-gates~\cite{alrahis21_OMLA}, we opt for MUX-based locking
which is considered more resilient~\cite{sisejkovic22}. Still, even MUX-based
schemes have been attacked recently~\cite{alrahis22_ML,alaql21}.
To render \TM\ resilient
against such advanced ML-based attacks, we devise the following implementation.

\textit{\ul{Locking Approach.}}
Unlike prior art,
\TM\
is \textit{not} based on obfuscating the netlist connectivity through MUX key-gates, but on using MUX key-gates for
regular locking.
Thus, for \TM, there is no information leakage arising from MUX key-gates for the connectivity and structure of the
design, which is the key vulnerability of prior art~\cite{alrahis22_ML}.
Instead, \TM\ employs simple but resilient key-gate structures with localized connectivity, described next.

\textit{\ul{Learning-Resilient Key-Gate Structures.}}
The design of \TM\ ensures that key-bits are fully randomized, without any correlation to each of the locked gate's
true functionality.
To do so, \TM\ instances render the locked gates interchangeable with respect to their complementary counterparts,
   e.g., a NAND gate can act as AND or NAND, only depending on the key-bit.

When locking gates using \TM\ instances, one picks randomly
from the different possible configurations (Fig.~\ref{fig:TM:gates}).
It is essential to note how the different configurations are structurally indistinguishable; only the
key-bits determine the true functionality. Also, key-bits are easy to randomize for any particular true functionality,
simply by randomly picking one of the possible configurations.
For simple complementary gates, like (a) AND and (b) NAND,
there are four different pairwise configurations which are indistinguishable on their own.
	Note how half the configurations are based on first transforming the original gates to their
	counterparts; this serves for
	further obfuscation of the overall layout regarding the distribution of gate types.

\textit{\ul{Locking of Complex Gates.}}
Concerning flip-flops (FFs), these can be locked as is, by connecting 
the Q and QN ports to a dedicated key-gate structure (Fig.~\ref{fig:TM:gates}(c)).\footnote{%
We note that, independently of our work, this key-gate structure was proposed by Karmakar et
al.~\cite{karmakar20}, though in the context of locking scan chains.}
For commercial libraries, where FF outputs are typically optimized and trimmed,
similar key-gate structures as for simple gates can be initiated (Fig.~\ref{fig:TM:gates}(d)).

For other complex gates like AOI,
we observe that complementary counterparts appear rarely in an optimized layout, if they are available at all in the library.
Thus, we defer locking of such complex gates as follows. When such complex gate is selected for locking, we search
its fan-out in depth-first manner for simple gates and FFs, covering all the fan-out paths.
Then, all the found simple gates and FFs are locked instead, which
essentially translates to locking the downstream structure of that complex gate.

\begin{figure}[tb]
\centering
\includegraphics[width=0.7\columnwidth]{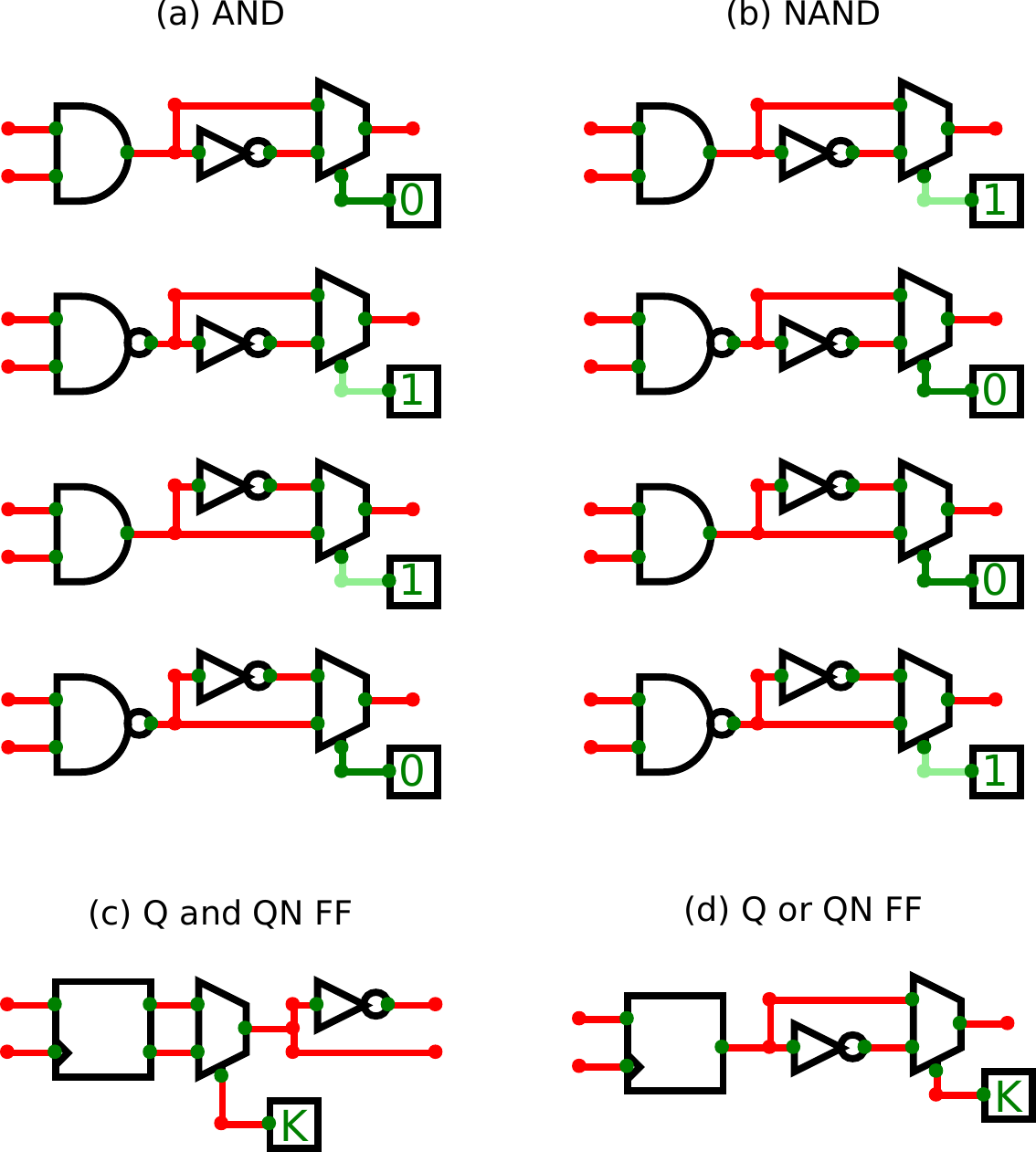}
\caption{Design of different \TM\ instances.
	The key-bit is connected to the MUX select line.
	(a, b) Locking of complementary (N)AND gates; other pairings of simple gates are locked similarly.
	(c, d) Locking of flip-flops (FFs) with different output configurations.
}
\label{fig:TM:gates}
\end{figure}

\textit{\ul{No Information Leakage from Physical Layout.}}
As indicated, there are \textit{no} direct correlations between the types of original versus the locked gate,
the connectivity
within and across \TM\ instances, and the
correct key-bit;
all are interchangeable, randomly chosen, and thus indistinguishable for an attacker.

It is also important to note that the two internal \TM\ nets connecting to the MUX inputs (or driving the MUX
fan-out for the case of Q and QN FFs, Fig.~\ref{fig:TM:gates}(c)) share a path and are, thus, both optimized/impacted
at once by EDA tools. Accordingly, an attacker seeking to study the underlying timing paths, driver
strengths, etc., would not gain any additional information.

\begin{figure}[tb]
	\centering
	\includegraphics[width=.9\columnwidth]{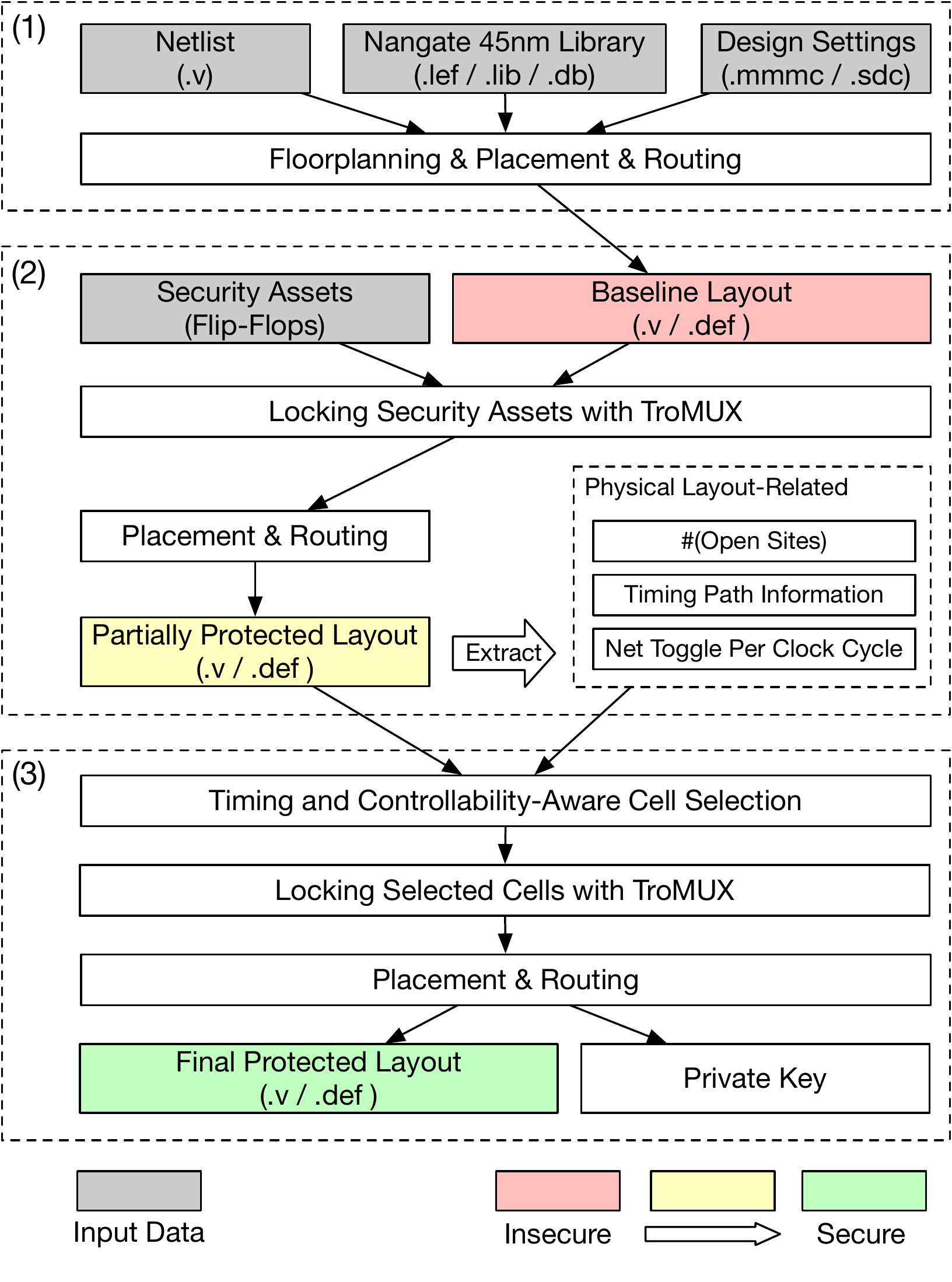}
\smallerspace
	\caption{Our physical-synthesis flow, which consists of three parts: (1) initial synthesis, (2) locking of
		security assets, and (3) locking considering timing and controllability.}
\label{fig:physical_synthesis_overall_flow}
\belowfigurespace
\end{figure}

\subsection{Physical Synthesis}

With the proposed locking scheme, we introduce a physical-synthesis flow that is capable of hardening any post-route
layout with negligible timing and area overheads. As outlined in Fig.~\ref{fig:physical_synthesis_overall_flow}, we
start with the original netlist and other necessary data to generate a baseline layout and then carefully interleave logic
locking and physical synthesis to produce a final protected layout.

\subsubsection{Two-Stage Locking Scheme}

In the first stage, we lock all security-critical FFs, as defined by assets.
After locking, we conduct placement and routing (P\&R), and obtain a partially protected layout (PPL). A PPL
has only FF assets protected; no other parts like LCNs and LCCs (i.e., the cells driving LCNs; both are targets for
		Trojan triggers) are protected yet.
Further, a PPL has relatively low utilization, leaving placement sites exposed to Trojan insertion.

In the second stage, we thus aim at locking as many cells as possible to fill the layout, also prioritizing the
locking of LCCs, all without degrading the design quality.
With the physical layout-related information obtained from the PPL, we first derive the number of cells to be locked,
then perform cell selection considering timing and controllability (detailed in Sec.~\ref{sec:cells_sel}).
After locking all selected cells, P\&R is re-run, and we finally retrieve a highly-utilized layout with all security-critical
components well protected through both \TM\ locking and much fewer placement sites left open.

\subsubsection{Storage of Key-Bits}
For each \TM\ instance, there will be one corresponding key-bit, requiring some facility to store all
key-bits.
We employ a large shift register for the following reasons:
\begin{itemize}
\item Prior art often considers adding a dedicated PI for each key-bit, which is not scalable. For us, we only require
two additional PIs (data in,
		load) for any number of key-bits.
\item Ours incurs negligible impact on performance and power,\footnote{%
For a design locked with $k$ key-bits, loading will take $k$ clock cycles.
This is done during initial boot-up, when the main circuitry is still hold in reset; runtime cost for such one-time
initialization are considered negligible.
Once the load signal signal is set low, the key-bits will remain stable,
consuming only some static FF power.
}
while the impact on area is even desired (see below).
\item The FFs used to build up the shift register are also helpful for
locally filling open placement sites as needed, whereas bulky memory blocks would rather complicate this task.
\item Finally, memory blocks are unavailable for the library used in this work, i.e., \textit{Nangate 45nm Open Cell
		Library}~\cite{NG45}. We require this library for the recently introduced ISPD'22 benchmarks for
		security closure of physical layouts~\cite{knechtel22_SCPL_ISPD}.
\end{itemize}

\subsubsection{On-Demand Key Length}
\TM\ instances occupy open placement sites with their INVs, MUXes, and FFs. 
After locking security assets,
we determine the additional key length
required to fill the physical layout at hand as follows:
\begin{align}
	k = floor\left(\frac{num\_open\_sites}{size(INV) + size(MUX) + size(FF) + \alpha}\right)
	\label{eq:keylength}
\end{align}
where $size(INV)$ represents the size of the smallest INV cell in the library, etc.,
and $\alpha$ is a parameter for timing budget.
As commercial tools will usually conduct timing optimization by gate sizing and buffer/inverter insertion, we need to reserve some additional space during locking. 
Accordingly, for designs where timing closure is more challenging, we will need a larger $\alpha$, and vice versa.

\input{incl/tables/table_notations.tex}

\subsubsection{Cell Selection Considering Timing and Controllability}
\label{sec:cells_sel}
Recall that, in the second locking stage, we lock selected cells to reduce the number of open sites and protect more
LCNs/LCCs. 
As key-gate structures can introduce further cell delays to related timing paths, the selection of cells to lock becomes critical for timing closure. 
Thus, we propose a scoring function $cellScore(c, N, P)$ to comprehensively describe the priority of a cell $c$ as
follows.
\begin{align}
	cellScore(c, N, P) = &\sum_{n \in N(c)} netScore(n, P), \\
	netScore(n, P) = &\frac{1}{1 + exp(-2\cdot MS(n, P))} \cdot \frac{1}{TPC(n) + 10^{-3}}, \\
	MS(n, P) = & 
		\begin{cases}
		\min\limits_{p \in P(n)} p.slack &, \text{if } |P(n)| > 0,\\
		-0.5 &, \text{otherwise},
		\end{cases}
\end{align}
with terms described in Table~\ref{tab:notations}. 
The cell score is represented as the sum of net scores to generalize to multi-output cases, e.g., both Q and QN of a FF
are used.
Further, we devise $netScore(n, P)$ to jointly consider timing and controllability. 
Using sigmoid, the score value remains positive even for negative but small slacks.
Besides, since $TPC(n) \in [0,2]$ and nets with $TPC \leq 0.1$ are considered as LCNs in this work,
we add a small margin ($10^{-3}$) to avoid both div-by-0 and $TPC(n)$ from dominating the score.
When calculating MS, some nets may not be covered by any reported timing paths (due to limitations of the commercial
tool). For those nets, $MS(n,P)$ returns -0.5, a fall-back value close to average negative slacks observed;
this value serves to conservatively penalize cells with unknown timing.

\input{incl/alg/net_selection.tex}

Using the above scoring, we propose an iterative cell-selection algorithm in Algorithm~\ref{algo:cell_selection}. 
In each iteration, we calculate cell scores (line 3--4) and pick those cells with the highest score (line 5--7). 
Next, we update the slacks of affected timing paths based on $\sigma$, a pessimistic estimate of delay introduced by
locking, defined as the sum of worst-case delays for INV\_X1 and MUX2\_X1 (i.e., the default \TM\ cells for our
		experiments), as derived from the libraries for the matching corner cases (line 8--10). 

Our initial experiments show that
the proposed two-stage locking scheme is more timing-friendly over a single-stage approach.
That is, when we did directly use the physical information extracted from the baseline layouts (marked in red in
Fig.~\ref{fig:physical_synthesis_overall_flow}) for cell selection and related locking,
it was much harder to achieve timing closure, due to the accumulation of slack estimation errors.

\section{Experimental Results}
\label{sec:results}

\subsection{Setup}

\subsubsection{Tools}

As indicated, we base our work on a commercial-grade design flow.
Without loss of generality, we use \textit{Cadence Innovus 20.14}
for physical synthesis.
The methodology is implemented in custom \textit{TCL} scripts and \textit{Python} code.
For security analysis using \textit{SCOPE} and \textit{MuxLink}, setup details are provided in 
Sec.~\ref{sec:experiments:security:ML}.

\subsubsection{Benchmarks}

We employ the benchmark suite from the ISPD'22 contest on security closure~\cite{knechtel22_SCPL_ISPD}.
The suite comprises a range of crypto cores as well as the \textit{openMSP430} microcontroller. 
As Table~\ref{tab:bm_stat} shows, the designs vary in terms
of complexity, utilization, size (cells, nets), available metal layers, timing constraints, and corners.
Since the suite was synthesized by legacy versions of \textit{Cadence Innovus}, we resynthesize all designs at our end
with floorplan utilization rates similar to the original post-route benchmark layouts; only
the rates for \textit{AES\_3}, \textit{openMSP430\_2}, and \textit{TDEA} are set 10\% lower, as needed to lock all
security assets.

\input{incl/tables/table_benchmark_stat.tex}
\input{incl/tables/table_exp1.tex}

\begin{figure*}[!t]
	\captionsetup[subfigure]{labelformat=empty}
		\centering
		\subfloat[]{\includegraphics[width=.32\linewidth]{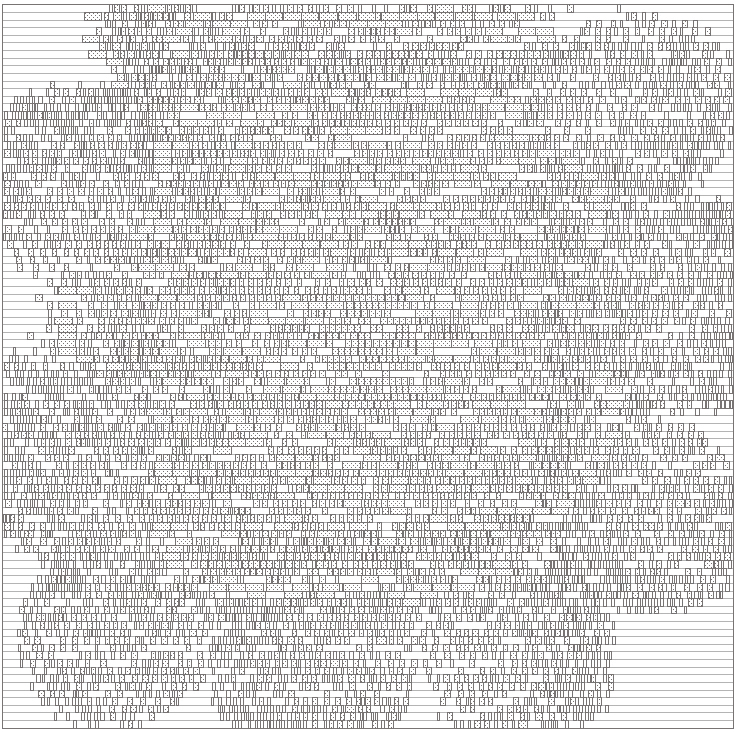}
		\label{fig:pre_logic_locking_demo} 		
		}
		\hfill
		\subfloat[]{\includegraphics[width=.32\linewidth]{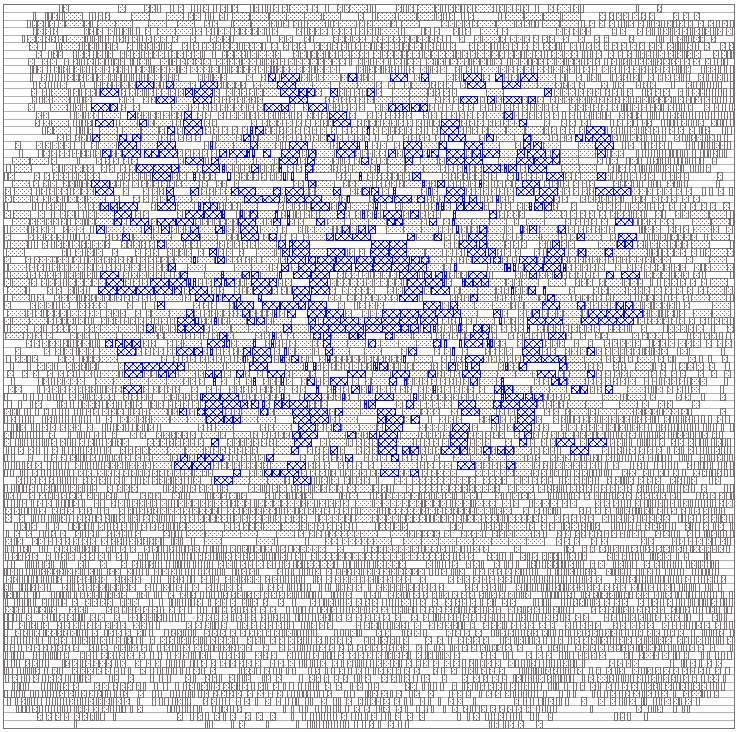}
	\label{fig:post_lock_asset_demo}
		}
		\hfill
		\subfloat[]{\includegraphics[width=.32\linewidth]{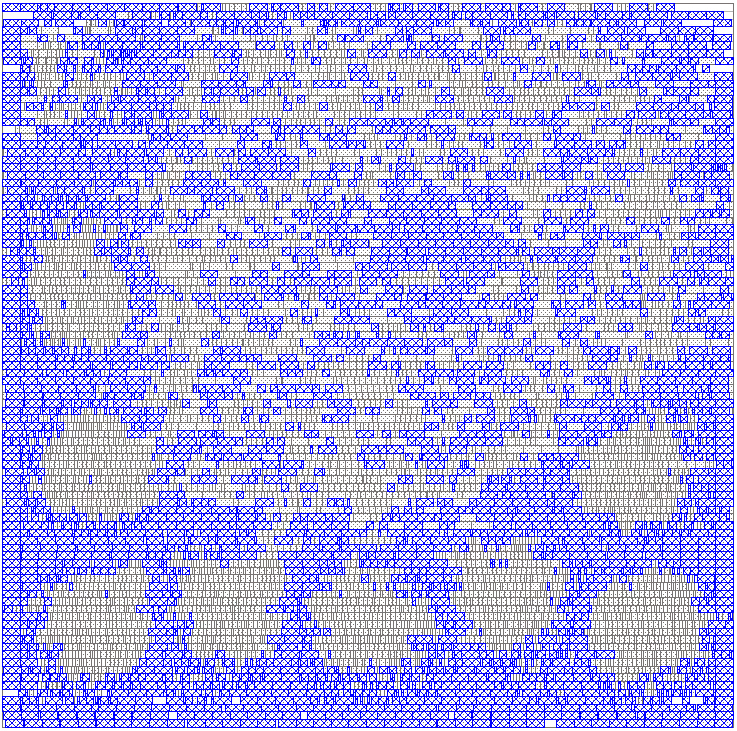}
	\label{fig:post_lock_lcn_demo}
		}
	\smallerspace
	\smallerspace
	\smallerspace
		\caption{Camellia layout, after initial synthesis (left), locking of security assets (middle), 
		and locking considering timing and controllability (right).
			The utilization increases from 49.5\% to 59.2\%, and eventually to 98.9\%. 
			Gates introduced by \TM\ instances are marked in blue while the other standard cells are filled with grey dots. 
		}
		\label{fig:logic_locking_demo}
		\belowfigurespace
\end{figure*}

\subsection{Results on the ISPD'22 Benchmark Suite}
We quantify security and layout results for the
resynthesized baseline layouts versus the final, protected layouts in Table~\ref{tab:exp1}. 

First, all the protected layouts show much higher utilization rates, thereby reducing open placement sites by 90.3\%
on average and rendering designs much more resilient against Trojan insertion in general.
Meanwhile, we achieve higher track utilization (defined as \textit{total routed wire length} / \textit{total track length}) compared with the baseline layout
such that routing from triggers to payloads will be more challenging for the attackers.
Second, as another layer of protection, recall that all security assets
are locked and that LCNs/LCCs are well locked in most layouts, except for those with relatively high initial
utilizations.
Third, all layouts are
	without any design rule check (DRC) violations, despite the ultra-high utilization.
This demonstrates the effectiveness of our proposed flow.
Fourth, total power is increased by 18.5\% on average, which seems reasonable given all the additional cells
introduced with \TM\ instances.

In Fig.~\ref{fig:logic_locking_demo}, we show the layouts of the exemplary \textit{Camellia} design in three stages as described in Fig.~\ref{fig:physical_synthesis_overall_flow}.
After locking assets, there are still a number of open sites. 
However, with the second locking stage, we manage to increase the utilization to as high as 98.9\%.

\subsection{ML-Based Attack Analysis}
\label{sec:experiments:security:ML}
Recall that we use logic locking to both protect security-critical components in particular and the layout in
general.
Attackers would want to undermine our locking and remove \TM\ instances, to be able to insert Trojans targeted on
assets and LCNs/LCCs in particular (Sec.~\ref{sec:tm}).
Thus,
we evaluate ours against state-of-the-art ML-based attacks \textit{SCOPE}~\cite{alaql21} and \textit{MuxLink}~\cite{alrahis22_ML}.

\textit{\ul{Setup for SCOPE.}} The attack expects designs in \textit{bench} format.
Thus, we resynthesize our locked benchmarks using the \textit{bench}-specific, limited set of cells,
	(i.e., AND, OR, NAND, NOR, INV, BUF, DFF, XOR, and XNOR)
	and convert the netlists into \textit{bench} format using an in-house \textit{Python} script.
	Since the attack cannot handle loops in the locked
	designs, we represent FFs as pseudo PIs/POs. Further, we use the default margin value of 0~\cite{alaql21}.

\textit{\ul{Setup for MuxLink.}} The original implementation supports only selected gates (i.e., AND, NAND, OR,
		NOR, INV, XOR, XNOR, BUF).
Since we use the ISPD'22 benchmarks where all gates from the
\textit{Nangate} library are used,
we extend the one-hot feature vectors of \textit{MuxLink} accordingly.
Consistent with the original operation of \textit{MuxLink}, we treat each pin of each gate as a node and the
connections between pins, including those going through gates, as edges. The connections between the input/output pins
of \TM\ key-gates are kept as test set for prediction, while all others (except PI or PO connections)
are used as training set. We adopt the same graph neural network configuration and training
hyperparameters as in~\cite{alrahis22_ML}; in particular, number of hops for extracting subgraphs is set to 3 and
the threshold in post-processing is set to 0.01.

\input{incl/tables/table_ML_attacks.tex}

\textit{\ul{Evaluation Metrics.}}
We report the performance of the
\textit{SCOPE} and \textit{MuxLink} attacks using the following established metrics: \textit{accuracy} (AC), \textit{precision} (PC),
\textit{key prediction accuracy} (KPA), and \textit{COnstant Propagation Effect} (COPE, \cite{alaql21});
all metrics are in percentage. AC measures the percentage of correctly deciphered
key-bits, i.e., $(k_{correct}/k_{total})$.
PC measures the 
correctly deciphered key-bits, optimistically considering every $X$/undeciphered value as a correct guess, i.e., $((k_{correct} +
k_{X})/k_{total})$.
KPA measures the correctly deciphered key-bits over the
entire prediction set, i.e., $(k_{correct}/(k_{total}-k_{X}))$.
COPE measures the vulnerability against the \textit{SCOPE} attack; COPE = 0\% means the attack fails entirely.

\textit{\ul{Results for SCOPE.}}
The results in Table~\ref{tab:ml_attack_result} show that
19.35\%/19.25\% of the key bits are correctly recovered on average. The average value of 0.065\% for COPE means that the attack fails
almost entirely. Further, the
average KPA of 53.72\% versus 46.28\% for the two keys\footnote{%
\textit{SCOPE} predicts the two mutually complementary keys, representing the possible configurations of `0' and `1'
	versus `1' and `0' for key-bits.}
indicates that \textit{SCOPE} is forced to random guessing for ours.
Note that the attack can only decipher a single bit for
\textit{TDEA}, thus resulting in KPA of 100\% and 0\% for that design.

\textit{\ul{Results for MuxLink.}}
The results in Table~\ref{tab:ml_attack_result} show that \textit{MuxLink}
can only predict, on average, 20.90\% of the key-bits. Moreover, the average KPA is 53.21\%, clearly indicating that
\textit{MuxLink} is forced to random guessing.
In contrast to prior art like \textit{D-MUX}~\cite{sisejkovic22,alrahis22_ML}---also showcased in
Table~\ref{tab:ml_attack_result}---ours is superior in thwarting the attack.

\subsection{Discussion of Prior Art}

Recall the review of related prior art in Sec.~\ref{sec:prior:locking} and the overview in
Table~\ref{tab:hl_comp}.
Note that none of the prior art released their results artifacts publicly. Further, most of the prior art uses different
technology libraries and implementation schemes, as well as different benchmarks. Thus, a direct comparison is
impractical.

Still, it is essential to recall the considerable limitations of
those studies and to forecast related implications.
For example,
for X(N)OR key-gates utilized by Samimi et al.~\cite{samimi16} and Marcelli et al.~\cite{marcelli17}, we note that Alrahis et
al.~\cite{alrahis21_OMLA} have independently shown that such locking can be circumvented with up to 97.22\% accuracy.
Thus, we argue that locking as applied by
Samimi et al.~\cite{samimi16} and Marcelli et al.~\cite{marcelli17} can be easily circumvented, unlike ours
(Sec.~\ref{sec:experiments:security:ML}).

\section{Conclusion}
\label{sec:conc}

We propose a MUX-based locking scheme, called \TM, along with a carefully tuned physical-synthesis flow, for preventing targeted Trojan insertion. 
To the best of our knowledge, ours is the first to systematically employ locking and layout-level means in unison for
Trojan prevention.
Results on the ISPD'22 contest benchmarks show that ours can reduce the number of open placement
sites by 90.3\% on average, with security-critical components secured by logic locking. 
Further, we demonstrate the superior resilience of ours against ML-based attacks:
on average, \textit{SCOPE} and \textit{MuxLink} can only correctly predict  
19.35\%/19.25\% and 20.90\%
of the key-bits, respectively,
with around 50\% 
key-prediction accuracy for both, indicating our scheme is enforcing random guessing.

\balance

\input{main.bbl}


\end{document}

%% file: abstract.tex
Due to cost benefits, supply chains of integrated circuits (ICs) are largely outsourced nowadays.
However, passing ICs through various third-party providers gives rise to many threats, like piracy of IC intellectual
property or insertion of hardware Trojans, i.e., malicious circuit modifications.

In this work, we proactively and systematically harden the physical layouts
of ICs
against post-design insertion of
Trojans.
Toward that end, we propose a multiplexer-based logic-locking scheme that is (i)~devised for layout-level
Trojan prevention,
(ii)~resilient against state-of-the-art, oracle-less machine learning attacks, and (iii)~fully integrated into a
tailored, yet generic, commercial-grade design flow.
Our work provides in-depth security and layout analysis on a challenging benchmark suite.
We show that ours can render layouts resilient, with reasonable overheads, against Trojan insertion in general and
also against
second-order attacks (i.e., adversaries seeking to bypass the locking defense in an oracle-less setting).

We release our layout artifacts for independent verification
\cite{release}
and we will release our methodology's source code.

%% file: incl/tables/table_prior.tex
\begin{table}
\setlength\tabcolsep{2.0pt}
	\small
	\caption{High-Level Comparison with Prior Art}
	\label{tab:hl_comp}
	\smallerspace
	\begin{tabular}{cccccc}
		\toprule
		\multirow{2}{*}{Work} & \multirow{2}{*}{Approach} & \multirow{2}{*}{Robust} &
		{Effective$^a$} & \multirow{2}{*}{Efficient} & Artifacts\\[1mm]
		& & & Targ. / Untarg. & & Available\\[1mm]
		\toprule
		\cite{dupuis14} & Locking & N~\cite{dupuis14,samimi16} & N$^{a'}$ / N & (Y) & N \\
		\midrule
		\cite{samimi16,marcelli17} & Locking & (N)$^b$ & (N)$^{a'}$ / N & (Y) & N \\
		\midrule
		\cite{sisejkovic19} & Locking & (?) & (?)$^{a'}$ / N & (Y) & N \\
		\midrule
		\cite{xiao14,ba15,ba16} & Addit.\ Logic & (?) & N / (?)$^{a'}$ & (Y) & N \\
		\midrule
		{\cite{hosseintalaee17,knechtel21_SCPL_ICCAD}} & Phys. Synth. & {(N)$^c$} &
		{N / (N)$^{a'}$} & {Y} & {N} \\
		\midrule
		\textbf{Ours:} & \textbf{Locking and} & \multirow{2}{*}{\textbf{Y}} &
		\multirow{2}{*}{\textbf{Y / (Y)}} & \multirow{2}{*}{\textbf{Y}}
		& \multirow{2}{*}{\textbf{Y~\cite{release}}} \\
		\textbf{\TM} & \textbf{Layout Filling} & & & & \\
		\bottomrule
	\end{tabular}\\[1mm]
	\footnotesize
	\begin{justify}
Notation: Y -- yes, N -- no, (?) -- unclear, (Y) -- yes but some caveat, (N) -- unlikely\\
	$^a$Two scenarios are differentiated: effective against targeted Trojans / against
	   untargeted Trojans. Robustness impacts effectiveness; related cases are labelled via ${a'}$.\\
	$^b$ The employed locking scheme has been broken in~\cite{alrahis21_OMLA}.\\
	$^c$We argue that such schemes can be reverted by adversaries.
	\end{justify}
\littlesmallerspace
\end{table}

%% file: incl/tables/table_notations.tex
\begin{table}[]
	\footnotesize
    \caption{Notations for Cell Selection}\label{tab:notations}
	\smallerspace
    \begin{tabular}{cc}
		\toprule
    {Term}   & {Description}                                 \\
		\toprule
    $C$      & The set of standard cell instances                   \\ \midrule
    $c$      & A cell instance from $C$                      \\ \midrule
    $N$      & The set of nets                             \\ \midrule
    $n$      & A net from $N$                                \\ \midrule
    $N(c)$   & The set of nets driven by $c$                 \\ \midrule
    $P$      & The set of timing paths                     \\ \midrule
    $P(n)$   & The set of timing paths covering $n$          \\ \midrule
    $MS(n, P)$ & The minimum slack of paths covering $n$ \\ \midrule
    $TPC(n)$ & The number of toggles per clock cycle for $n$ \\
	    \bottomrule
    \end{tabular}
    \belowtablespace
\end{table}

%% file: incl/alg/net_selection.tex
\begin{algorithm}[]
	\footnotesize
	\caption{
		Cell Selection Considering Timing and Controllability}\label{algo:cell_selection}
	\begin{algorithmic}[1]
		\Require Standard Cells $ C $, Nets $ N $, Timing Paths $ P $, Key Length $K$, Locking Delay $\sigma$.
		\Ensure The Set of Cells to Lock $ C' $.
		\State $C' \gets \{\}$; 
		\While{ $|C'| < K $ }
			\ForAll{cell $c \in C$} 
				\State $c.score \gets cellScore(c, N, P) $; \Comment{Score calculation for each cell}
			\EndFor
			\State Let $c_{h}$ be the cell with the highest score in $C$;
			\State $C'.append(c_{h})$;
			\State $C.remove(c_{h})$;
			
			\ForAll{net $n \in N(c_{h})$} 
				\ForAll{timing path $p \in P(n)$}
					\State $p.slack \gets p.slack - \sigma$; \Comment{Pessimistic slack estimation} 
				\EndFor
			\EndFor
		\EndWhile
		
		\State \Return $C'$;
	\end{algorithmic}
\end{algorithm}

%% file: incl/tables/table_benchmark_stat.tex
\begin{table}[t]
	\centering
    \caption{Benchmark Statistics}\label{tab:bm_stat}
    \smallerspace

	\resizebox{.95\columnwidth}{!}{%
	\begin{tabular}{c|cccccc}
		\toprule
			Design        & F. Utils & \#(Cells) & \#(Nets) & \#(ML) & CP & Corner  \\
			\midrule
			AES\_1        & 75.0\%   & 16509     & 19694    & 10     & 1       & typical \\
			AES\_2        & 75.0\%   & 16509     & 19694    & 10     & 1       & typical \\
			AES\_3        & 85.0\%   & 15836     & 19020    & 10     & 1       & typical \\
			Camellia      & 50.0\%   & 6710      & 7160     & 6      & 10      & slow    \\
			CAST          & 50.0\%   & 12682     & 13057    & 6      & 10      & slow    \\
			MISTY         & 50.0\%   & 9517      & 9904     & 6      & 10      & slow    \\
			openMSP430\_1 & 50.0\%   & 4690      & 5312     & 6      & 30      & slow    \\
			openMSP430\_2 & 70.0\%   & 5921      & 6550     & 6      & 8       & slow    \\
			PRESENT       & 50.0\%   & 868       & 1046     & 6      & 10      & slow    \\
			SEED          & 50.0\%   & 12682     & 13057    & 6      & 10      & slow    \\
			SPARX         & 50.0\%   & 8146      & 10884    & 6      & 10      & slow    \\
			TDEA          & 70.0\%   & 2269      & 2594     & 6      & 4       & slow  \\ 
		\bottomrule 
	\end{tabular}
	}

	\begin{tablenotes}
		\footnotesize
		\item  * 
		F. Utils: floorplanning utilization for resynthesis; 
		\#(Cells/Nets): number of cells/nets in original netlists;
		\#(ML): number of metal layers;
		CP (ns): clock period;
		Corner: typical is characterized for 1.1V and 25C, slow for 0.95V and 125C.
	\end{tablenotes}
	\belowtablespace
\end{table}

%% file: incl/tables/table_exp1.tex
\begin{table*}[th]
    \centering
    \caption{Layout and Security Results for Ours on the ISPD'22 Contest Benchmark Suite}\label{tab:exp1}
    \smallerspace
    \resizebox{.95\linewidth}{!}{%

    \begin{tabular}{c|cccccc|cccccccccc}
        \toprule
        \multirow{2}{*}{Design} & \multicolumn{6}{c|}{Baseline Layout (Resynthesized)} & \multicolumn{10}{c}{Protected Layout (Final)}                                                                   \\ \cmidrule{2-17}        
        & Utils       & \#(Open)  & TU   & WNS        & TNS        & Power     & Utils   & \#(Open) & $\Delta$(Open) & TU & WNS    & TNS    & Power  & \#(LSA)/\#(SA) & \#(LLCC)/\#(LCC) & KL \\
        \midrule
        AES\_1        & 75.4\% & 43,980 & 8.7\% & 0.000  & 0.000  & 59.957 & 96.2\% & 6,838 & -84.5\% & 11.1\% & -0.013 & -0.043 & 60.552 & 291/291     & 884/1,389 & 1,199 \\
        AES\_2        & 75.1\% & 44,420 & 8.7\% & -0.001 & -0.003 & 59.441 & 96.3\% & 6,649 & -85.0\% & 10.9\% & -0.008 & -0.017 & 61.040 & 291/291     & 908/1,431 & 1,202 \\
        AES\_3        & 85.7\% & 22,129 & 9.7\% & -0.001 & -0.002 & 59.869 & 96.6\% & 5,225 & -76.4\% & 11.2\% & -0.031 & -4.657 & 62.787 & 291/291     & 150/1,310 & 441   \\
        Camellia      & 49.5\% & 33,919 & 9.5\% & 1.194  & 0.000  & 1.233  & 98.9\% & 753   & -97.8\% & 13.3\% & 0.015  & 0.000  & 1.488  & 256/256     & 724/724   & 1,271 \\
        CAST          & 49.1\% & 54,444 & 9.1\% & 0.047  & 0.000  & 3.136  & 93.6\% & 6,879 & -87.4\% & 12.7\% & -0.134 & -1.455 & 3.912  & 192/192     & 983/994   & 1,572 \\
        MISTY         & 48.5\% & 43,359 & 8.6\% & -0.021 & -0.037 & 2.238  & 94.4\% & 4,686 & -89.2\% & 12.4\% & -0.140 & -1.515 & 2.966  & 204/204     & 307/312   & 1,215 \\
        openMSP430\_1 & 49.7\% & 32,799 & 6.2\% & 0.000  & 0.000  & 0.375  & 97.9\% & 1,389 & -95.8\% & 12.6\% & 0.000  & 0.000  & 0.544  & 340/340     & 441/456   & 1,218 \\
        openMSP430\_2 & 70.5\% & 15,125 & 7.7\% & 0.000  & 0.000  & 1.239  & 97.1\% & 1,497 & -90.1\% & 10.6\% & -0.006 & -0.015 & 1.369  & 334/334     & 209/696   & 543   \\
        PRESENT       & 49.8\% & 6,284  & 4.4\% & 6.694  & 0.000  & 0.198  & 98.0\% & 245   & -96.1\% & 6.8\%  & 4.935  & 0.000  & 0.235  & 80/80       & 3/3       & 241   \\
        SEED          & 49.1\% & 54,444 & 9.1\% & 0.047  & 0.000  & 3.136  & 94.3\% & 6,056 & -88.9\% & 12.7\% & -0.182 & -1.072 & 3.908  & 195/195     & 979/989   & 1,612 \\
        SPARX         & 49.9\% & 69,979 & 6.3\% & 2.452  & 0.000  & 2.164  & 98.8\% & 1,658 & -97.6\% & 14.0\% & 0.027  & 0.000  & 2.822  & 2,176/2,176 & 361/375   & 2,582 \\
        TDEA          & 70.4\% & 5,559  & 6.8\% & 0.049  & 0.000  & 1.084  & 98.4\% & 304   & -94.5\% & 8.0\%  & 0.027  & 0.000  & 1.153  & 168/168     & 6/47      & 214  \\
    \bottomrule    
    \end{tabular}
    }

    \begin{tablenotes}
		\footnotesize
        \item  * Utils: utilization after physical synthesis;
		\#(Open): number of open sites;
        TU: track utilization;
        WNS (ns): worst negative slack;
		TNS (ns): total negative slack;
        Power (mW): total power;
        $\Delta$(Open): reduction of number of open sites;
        SA: security assets;
        LSA: locked security assets;
        LCC: low-controllable cells;
        LLCC: locked low-controllable cells;
        KL: key length (bits).

	\end{tablenotes}
\end{table*}

%% file: incl/tables/table_ML_attacks.tex
\begin{table*}[h]
    \centering
    \caption{ML-Based Attack Results on Different Locking Schemes}
    \label{tab:ml_attack_result}
    \smallerspace
    \resizebox{.9\linewidth}{!}{
        \begin{tabular}{cc|cccccc|cccccccc}
            \hline
            \multirow{3}{*}{Design} & \multirow{3}{*}{KL} & \multicolumn{6}{c|}{\textit{SCOPE}~\cite{alaql21}}                                                                                                                                   & \multicolumn{8}{c}{\textit{MuxLink}~\cite{alrahis22_ML}}                                                                      \\ \cline{3-16} 
                                    &                     & \multicolumn{1}{c|}{\multirow{2}{*}{COPE (\%)}} & \multicolumn{2}{c|}{\TM\ Key 1}       & \multicolumn{2}{c|}{\TM\ Key 2}       & \multirow{2}{*}{\#(X)} & \multicolumn{4}{c|}{\TM}                               & \multicolumn{4}{c}{\textit{D-MUX}~\cite{alrahis22_ML,sisejkovic22}}            \\ \cline{4-7} \cline{9-16} 
                                    &                     & \multicolumn{1}{c|}{}                           & AC (\%) & \multicolumn{1}{c|}{KPA (\%)} & AC (\%) & \multicolumn{1}{c|}{KPA (\%)} &                        & AC (\%) & PC (\%) & KPA (\%) & \multicolumn{1}{c|}{\#(X)} & AC (\%) & PC (\%) & KPA (\%) & \#(X) \\ \hline
            AES\_1                  & 1,199               & \multicolumn{1}{c|}{0.1919}                     & 28.86   & 48.39                         & 30.78   & \multicolumn{1}{c|}{51.61}    & 484                    & 28.77   & 71.48   & 50.22    & \multicolumn{1}{c|}{512}   & 78.15   & 79.90   & 79.54    & 21    \\
            AES\_2                  & 1,202               & \multicolumn{1}{c|}{0.1864}                     & 31.53   & 52.71                         & 28.29   & \multicolumn{1}{c|}{47.29}    & 483                    & 28.12   & 75.04   & 52.98    & \multicolumn{1}{c|}{564}   & 80.45   & 81.70   & 81.47    & 15    \\
            AES\_3                  & 441                 & \multicolumn{1}{c|}{0.1075}                     & 23.36   & 48.13                         & 25.17   & \multicolumn{1}{c|}{51.87}    & 227                    & 8.16    & 90.02   & 45.00    & \multicolumn{1}{c|}{361}   & 88.44   & 89.12   & 89.04    & 3     \\
            Camellia                & 1,271               & \multicolumn{1}{c|}{0.0240}                     & 14.16   & 51.72                         & 13.22   & \multicolumn{1}{c|}{48.28}    & 923                    & 26.12   & 73.80   & 49.92    & \multicolumn{1}{c|}{606}   & 90.64   & 92.60   & 92.46    & 25    \\
            CAST                    & 1,572               & \multicolumn{1}{c|}{0.0359}                     & 18.58   & 50.52                         & 18.19   & \multicolumn{1}{c|}{49.48}    & 994                    & 37.47   & 66.03   & 52.45    & \multicolumn{1}{c|}{449}   & 93.74   & 94.32   & 94.29    & 7     \\
            MISTY                   & 1,215               & \multicolumn{1}{c|}{0.1036}                     & 23.70   & 48.90                         & 24.77   & \multicolumn{1}{c|}{51.10}    & 626                    & 33.33   & 65.84   & 49.39    & \multicolumn{1}{c|}{395}   & 97.84   & 98.03   & 98.02    & 3     \\
            openMSP430\_1           & 1,218               & \multicolumn{1}{c|}{0.0432}                     & 19.87   & 49.90                         & 19.95   & \multicolumn{1}{c|}{50.10}    & 733                    & 23.56   & 76.60   & 50.17    & \multicolumn{1}{c|}{646}   & NA      & NA      & NA       & NA    \\
            openMSP430\_2           & 543                 & \multicolumn{1}{c|}{0.0306}                     & 29.28   & 52.13                         & 26.89   & \multicolumn{1}{c|}{47.87}    & 238                    & 9.39    & 90.42   & 49.51    & \multicolumn{1}{c|}{440}   & 66.85   & 69.98   & 69.01    & 17    \\
            PRESENT                 & 241                 & \multicolumn{1}{c|}{0.0434}                     & 2.49    & 42.86                         & 3.32    & \multicolumn{1}{c|}{57.14}    & 227                    & 11.62   & 95.02   & 70.00    & \multicolumn{1}{c|}{201}   & NA      & NA      & NA       & NA    \\
            SEED                    & 1,612               & \multicolumn{1}{c|}{0.0343}                     & 18.55   & 49.10                         & 19.23   & \multicolumn{1}{c|}{50.90}    & 1,003                   & 37.03   & 64.02   & 50.72    & \multicolumn{1}{c|}{435}   & 97.33   & 97.70   & 97.70    & 6     \\
            SPARX                   & 2,582               & \multicolumn{1}{c|}{0.0176}                     & 21.42   & 50.23                         & 21.22   & \multicolumn{1}{c|}{49.77}    & 1,481                   & 5.38    & 94.93   & 51.48    & \multicolumn{1}{c|}{2,312}  & NA      & NA      & NA       & NA    \\
            TDEA                    & 214                 & \multicolumn{1}{c|}{0.0001}                     & 0.47    & 100.00                        & 0.00    & \multicolumn{1}{c|}{0.00}     & 213                    & 1.87    & 99.07   & 66.67    & \multicolumn{1}{c|}{208}   & NA      & NA      & NA       & NA    \\ \hline
            Avg.                    & -                   & \multicolumn{1}{c|}{0.0682}                     & 19.35   & 53.72                         & 19.25   & \multicolumn{1}{c|}{46.28}    & -                      & 20.90   & 80.19   & 53.21    & \multicolumn{1}{c|}{-}     & 86.68   & 87.92   & 87.69    & -     \\ \hline
            \end{tabular}

    }

    \begin{tablenotes}
		\footnotesize
        \item  * AC: accuracy;
PC: precision;
KPA: key prediction\ accuracy;
COPE: COnstant Prop.\ Effect;
	\#(X): number of undeciphered key-bits;
NA: Locking using the script in~\cite{alrahis22_ML} fails.
\end{tablenotes}
\belowtablespace
\end{table*}

%% file: main.bbl

%% file: main.bbl
\begin{thebibliography}{33}


\ifx \showCODEN    \undefined \def \showCODEN     #1{\unskip}     \fi
\ifx \showDOI      \undefined \def \showDOI       #1{#1}\fi
\ifx \showISBNx    \undefined \def \showISBNx     #1{\unskip}     \fi
\ifx \showISBNxiii \undefined \def \showISBNxiii  #1{\unskip}     \fi
\ifx \showISSN     \undefined \def \showISSN      #1{\unskip}     \fi
\ifx \showLCCN     \undefined \def \showLCCN      #1{\unskip}     \fi
\ifx \shownote     \undefined \def \shownote      #1{#1}          \fi
\ifx \showarticletitle \undefined \def \showarticletitle #1{#1}   \fi
\ifx \showURL      \undefined \def \showURL       {\relax}        \fi
\providecommand\bibfield[2]{#2}
\providecommand\bibinfo[2]{#2}
\providecommand\natexlab[1]{#1}
\providecommand\showeprint[2][]{arXiv:#2}

\bibitem[Alaql et~al\mbox{.}(2021)]%
        {alaql21}
\bibfield{author}{\bibinfo{person}{A. Alaql} {et~al\mbox{.}}}
  \bibinfo{year}{2021}\natexlab{}.
\newblock \showarticletitle{SCOPE: Synthesis-Based Constant Propagation Attack
  on Logic Locking}.
\newblock \bibinfo{journal}{\emph{Trans. VLSI}} \bibinfo{volume}{29},
  \bibinfo{number}{8} (\bibinfo{year}{2021}), \bibinfo{pages}{1529--1542}.
\newblock


\bibitem[Alrahis et~al\mbox{.}(2022a)]%
        {alrahis22_ML}
\bibfield{author}{\bibinfo{person}{L. Alrahis} {et~al\mbox{.}}}
  \bibinfo{year}{2022}\natexlab{a}.
\newblock \showarticletitle{MuxLink: Circumventing Learning-Resilient
  MUX-Locking Using Graph Neural Network-based Link Prediction}. In
  \bibinfo{booktitle}{\emph{Proc. DATE}}. \bibinfo{pages}{694--699}.
\newblock


\bibitem[Alrahis et~al\mbox{.}(2022b)]%
        {alrahis21_OMLA}
\bibfield{author}{\bibinfo{person}{L. Alrahis} {et~al\mbox{.}}}
  \bibinfo{year}{2022}\natexlab{b}.
\newblock \showarticletitle{OMLA: An Oracle-Less Machine Learning-Based Attack
  on Logic Locking}.
\newblock \bibinfo{journal}{\emph{TCS}} \bibinfo{volume}{69},
  \bibinfo{number}{3} (\bibinfo{year}{2022}), \bibinfo{pages}{1602--1606}.
\newblock


\bibitem[Ba et~al\mbox{.}(2015)]%
        {ba15}
\bibfield{author}{\bibinfo{person}{P.-S. Ba} {et~al\mbox{.}}}
  \bibinfo{year}{2015}\natexlab{}.
\newblock \showarticletitle{Hardware Trojan prevention using layout-level
  design approach}. In \bibinfo{booktitle}{\emph{Proc. Europ, Conf. Circ.
  Theory Des.}} \bibinfo{pages}{1--4}.
\newblock


\bibitem[Ba et~al\mbox{.}(2016)]%
        {ba16}
\bibfield{author}{\bibinfo{person}{P.-S. Ba} {et~al\mbox{.}}}
  \bibinfo{year}{2016}\natexlab{}.
\newblock \showarticletitle{Hardware Trust through Layout Filling: A Hardware
  Trojan Prevention Technique}. In \bibinfo{booktitle}{\emph{Proc. ISVLSI}}.
  \bibinfo{pages}{254--259}.
\newblock


\bibitem[Chakraborty et~al\mbox{.}(2018)]%
        {chakraborty18}
\bibfield{author}{\bibinfo{person}{P. Chakraborty} {et~al\mbox{.}}}
  \bibinfo{year}{2018}\natexlab{}.
\newblock \showarticletitle{SAIL: Machine Learning Guided Structural Analysis
  Attack on Hardware Obfuscation}. In \bibinfo{booktitle}{\emph{Proc. AHOST}}.
  \bibinfo{pages}{56--61}.
\newblock


\bibitem[Chen et~al\mbox{.}(2018)]%
        {chen18}
\bibfield{author}{\bibinfo{person}{X. Chen} {et~al\mbox{.}}}
  \bibinfo{year}{2018}\natexlab{}.
\newblock \showarticletitle{Hardware Trojan Detection in Third-Party Digital
  Intellectual Property Cores by Multilevel Feature Analysis}.
\newblock \bibinfo{journal}{\emph{TCAD}} \bibinfo{volume}{37},
  \bibinfo{number}{7} (\bibinfo{year}{2018}), \bibinfo{pages}{1370--1383}.
\newblock
\showISSN{0278-0070}


\bibitem[Deng et~al\mbox{.}(2020)]%
        {deng20}
\bibfield{author}{\bibinfo{person}{D. Deng} {et~al\mbox{.}}}
  \bibinfo{year}{2020}\natexlab{}.
\newblock \showarticletitle{Novel Design Strategy Toward A2 Trojan Detection
  Based on Built-In Acceleration Structure}.
\newblock \bibinfo{journal}{\emph{TCAD}} \bibinfo{volume}{39},
  \bibinfo{number}{12} (\bibinfo{year}{2020}), \bibinfo{pages}{4496--4509}.
\newblock


\bibitem[Dong et~al\mbox{.}(2020)]%
        {dong20}
\bibfield{author}{\bibinfo{person}{C. Dong} {et~al\mbox{.}}}
  \bibinfo{year}{2020}\natexlab{}.
\newblock \showarticletitle{Hardware Trojans in Chips: A Survey for Detection
  and Prevention}.
\newblock \bibinfo{journal}{\emph{Sensors}} \bibinfo{volume}{20},
  \bibinfo{number}{18} (\bibinfo{year}{2020}).
\newblock


\bibitem[Dupuis et~al\mbox{.}(2014)]%
        {dupuis14}
\bibfield{author}{\bibinfo{person}{S. Dupuis} {et~al\mbox{.}}}
  \bibinfo{year}{2014}\natexlab{}.
\newblock \showarticletitle{A novel hardware logic encryption technique for
  thwarting illegal overproduction and Hardware Trojans}. In
  \bibinfo{booktitle}{\emph{Proc. IOLTS}}. \bibinfo{pages}{49--54}.
\newblock


\bibitem[Guimar{\~a}es et~al\mbox{.}(2017)]%
        {guimaraes17}
\bibfield{author}{\bibinfo{person}{L.~A. Guimar{\~a}es} {et~al\mbox{.}}}
  \bibinfo{year}{2017}\natexlab{}.
\newblock \showarticletitle{Detection of Layout-Level Trojans by Monitoring
  Substrate with Preexisting Built-in Sensors}. In
  \bibinfo{booktitle}{\emph{Proc. ISVLSI}}. \bibinfo{pages}{290--295}.
\newblock


\bibitem[Guo et~al\mbox{.}(2019)]%
        {guo19}
\bibfield{author}{\bibinfo{person}{X. Guo} {et~al\mbox{.}}}
  \bibinfo{year}{2019}\natexlab{}.
\newblock \showarticletitle{When Capacitors Attack: Formal Method Driven Design
  and Detection of Charge-Domain Trojans}. In \bibinfo{booktitle}{\emph{Proc.
  DATE}}. \bibinfo{pages}{1727--1732}.
\newblock


\bibitem[Hossein-Talaee and Jahanian(2017)]%
        {hosseintalaee17}
\bibfield{author}{\bibinfo{person}{H. Hossein-Talaee} {and} \bibinfo{person}{A.
  Jahanian}.} \bibinfo{year}{2017}\natexlab{}.
\newblock \showarticletitle{Layout Vulnerability Reduction against Trojan
  Insertion Using Security-Aware White Space Distribution}. In
  \bibinfo{booktitle}{\emph{Proc. ISVLSI}}.
\newblock


\bibitem[Hou et~al\mbox{.}(2018)]%
        {hou18}
\bibfield{author}{\bibinfo{person}{Y. Hou} {et~al\mbox{.}}}
  \bibinfo{year}{2018}\natexlab{}.
\newblock \showarticletitle{R2D2: Runtime reassurance and detection of A2
  Trojan}. In \bibinfo{booktitle}{\emph{Proc. HOST}}.
  \bibinfo{pages}{195--200}.
\newblock


\bibitem[{Hu} et~al\mbox{.}(2020)]%
        {wu21}
\bibfield{author}{\bibinfo{person}{W. {Hu}} {et~al\mbox{.}}}
  \bibinfo{year}{2020}\natexlab{}.
\newblock \showarticletitle{An Overview of Hardware Security and Trust:
  Threats, Countermeasures and Design Tools}.
\newblock \bibinfo{journal}{\emph{TCAD}} (\bibinfo{year}{2020}).
\newblock


\bibitem[Karmakar and Chattopadhyay(2020)]%
        {karmakar20}
\bibfield{author}{\bibinfo{person}{R. Karmakar} {and} \bibinfo{person}{S.
  Chattopadhyay}.} \bibinfo{year}{2020}\natexlab{}.
\newblock \showarticletitle{On Securing Scan Obfuscation Strategies Against
  ScanSAT Attack}. In \bibinfo{booktitle}{\emph{Proc. ISQED}}.
  \bibinfo{pages}{213--218}.
\newblock


\bibitem[Knechtel et~al\mbox{.}(2021)]%
        {knechtel21_SCPL_ICCAD}
\bibfield{author}{\bibinfo{person}{J. Knechtel} {et~al\mbox{.}}}
  \bibinfo{year}{2021}\natexlab{}.
\newblock \showarticletitle{Security Closure of Physical Layouts}. In
  \bibinfo{booktitle}{\emph{Proc. ICCAD}}. \bibinfo{pages}{1--9}.
\newblock


\bibitem[Knechtel et~al\mbox{.}(2022)]%
        {knechtel22_SCPL_ISPD}
\bibfield{author}{\bibinfo{person}{J. Knechtel} {et~al\mbox{.}}}
  \bibinfo{year}{2022}\natexlab{}.
\newblock \showarticletitle{Benchmarking Security Closure of Physical Layouts:
  ISPD 2022 Contest}. In \bibinfo{booktitle}{\emph{Proc. ISPD}}.
  \bibinfo{pages}{221--228}.
\newblock


\bibitem[Marcelli et~al\mbox{.}(2017)]%
        {marcelli17}
\bibfield{author}{\bibinfo{person}{A. Marcelli} {et~al\mbox{.}}}
  \bibinfo{year}{2017}\natexlab{}.
\newblock \showarticletitle{An evolutionary approach to hardware encryption and
  Trojan-horse mitigation}. In \bibinfo{booktitle}{\emph{Proc. DATE}}.
  \bibinfo{pages}{1593--1598}.
\newblock


\bibitem[Nangate(2011)]%
        {NG45}
Nangate Inc \bibinfo{year}{2011}\natexlab{}.
\newblock \bibinfo{booktitle}{\emph{{NanGate FreePDK45 Open Cell Library}}}.
\newblock Nangate Inc.
\newblock
\urldef\tempurl%
\url{http://www.nangate.com/?page_id=2325}
\showURL{%
\tempurl}


\bibitem[Ravi et~al\mbox{.}(2019)]%
        {ravi19}
\bibfield{author}{\bibinfo{person}{P. Ravi} {et~al\mbox{.}}}
  \bibinfo{year}{2019}\natexlab{}.
\newblock \showarticletitle{Security is an Architectural Design Constraint}.
\newblock \bibinfo{journal}{\emph{Microprocess. Microsyst.}}
  \bibinfo{volume}{68}, \bibinfo{number}{C} (\bibinfo{year}{2019}),
  \bibinfo{pages}{17--27}.
\newblock


\bibitem[Samimi et~al\mbox{.}(2016)]%
        {samimi16}
\bibfield{author}{\bibinfo{person}{M.~S. Samimi} {et~al\mbox{.}}}
  \bibinfo{year}{2016}\natexlab{}.
\newblock \showarticletitle{Hardware enlightening: No where to hide your
  Hardware Trojans!}. In \bibinfo{booktitle}{\emph{Proc. IOLTS}}.
  \bibinfo{pages}{251--256}.
\newblock


\bibitem[Sigl(2011)]%
        {sigl11}
\bibfield{author}{\bibinfo{person}{G. Sigl}.} \bibinfo{year}{2011}\natexlab{}.
\newblock \showarticletitle{Keynote address: Design of secure systems -- Where
  are the EDA tools?}. In \bibinfo{booktitle}{\emph{Proc. ICCAD}}.
  \bibinfo{pages}{1--1}.
\newblock


\bibitem[Trippel et~al\mbox{.}(2020)]%
        {trippel20}
\bibfield{author}{\bibinfo{person}{T. Trippel} {et~al\mbox{.}}}
  \bibinfo{year}{2020}\natexlab{}.
\newblock \showarticletitle{ICAS: an Extensible Framework for Estimating the
  Susceptibility of IC Layouts to Additive Trojans}. In
  \bibinfo{booktitle}{\emph{Proc. SP}}. \bibinfo{pages}{1742--1759}.
\newblock


\bibitem[Vashistha et~al\mbox{.}(2018)]%
        {vashistha18}
\bibfield{author}{\bibinfo{person}{N. Vashistha} {et~al\mbox{.}}}
  \bibinfo{year}{2018}\natexlab{}.
\newblock \showarticletitle{Trojan Scanner: Detecting Hardware Trojans with
  Rapid {SEM} Imaging combined with Image Processing and Machine Learning}. In
  \bibinfo{booktitle}{\emph{ISTFA}}.
\newblock


\bibitem[Vijayan et~al\mbox{.}(2020)]%
        {vijayan20}
\bibfield{author}{\bibinfo{person}{A. Vijayan} {et~al\mbox{.}}}
  \bibinfo{year}{2020}\natexlab{}.
\newblock \showarticletitle{Runtime Identification of Hardware Trojans by
  Feature Analysis on Gate-Level Unstructured Data and Anomaly Detection}.
\newblock \bibinfo{journal}{\emph{TODAES}} \bibinfo{volume}{25},
  \bibinfo{number}{4} (\bibinfo{year}{2020}).
\newblock


\bibitem[\v{S}i\v{s}ejkovi\'{c} et~al\mbox{.}(2019)]%
        {sisejkovic19}
\bibfield{author}{\bibinfo{person}{D. \v{S}i\v{s}ejkovi\'{c}} {et~al\mbox{.}}}
  \bibinfo{year}{2019}\natexlab{}.
\newblock \showarticletitle{Control-Lock: Securing Processor Cores Against
  Software-Controlled Hardware Trojans}. In \bibinfo{booktitle}{\emph{Proc.
  GLSVLSI}}. \bibinfo{pages}{27--32}.
\newblock


\bibitem[\v{S}i\v{s}ejkovi\'{c} et~al\mbox{.}(2022)]%
        {sisejkovic22}
\bibfield{author}{\bibinfo{person}{D. \v{S}i\v{s}ejkovi\'{c}},
  \bibinfo{person}{F. Merchant}, \bibinfo{person}{L.~M. Reimann}, {and}
  \bibinfo{person}{R. Leupers}.} \bibinfo{year}{2022}\natexlab{}.
\newblock \showarticletitle{Deceptive Logic Locking for Hardware Integrity
  Protection Against Machine Learning Attacks}.
\newblock \bibinfo{journal}{\emph{TCAD}} \bibinfo{volume}{41},
  \bibinfo{number}{6} (\bibinfo{year}{2022}), \bibinfo{pages}{1716--1729}.
\newblock


\bibitem[Wang et~al\mbox{.}(2022)]%
        {release}
\bibfield{author}{\bibinfo{person}{F. Wang} {et~al\mbox{.}}}
  \bibinfo{year}{2022}\natexlab{}.
\newblock \bibinfo{booktitle}{\emph{Baseline and Protected Layouts}}.
\newblock
\urldef\tempurl%
\url{https://drive.google.com/drive/folders/1A_Cy6w2n31_wuPKVayz50R-1lfXfC4vH?usp=sharing}
\showURL{%
\tempurl}


\bibitem[Xiao et~al\mbox{.}(2014)]%
        {xiao14}
\bibfield{author}{\bibinfo{person}{K. Xiao} {et~al\mbox{.}}}
  \bibinfo{year}{2014}\natexlab{}.
\newblock \showarticletitle{A Novel Built-In Self-Authentication Technique to
  Prevent Inserting Hardware Trojans}.
\newblock \bibinfo{journal}{\emph{TCAD}} \bibinfo{volume}{33},
  \bibinfo{number}{12} (\bibinfo{year}{2014}), \bibinfo{pages}{1778--1791}.
\newblock
\showISSN{0278-0070}


\bibitem[Xiao et~al\mbox{.}(2016)]%
        {xiao16}
\bibfield{author}{\bibinfo{person}{K. Xiao} {et~al\mbox{.}}}
  \bibinfo{year}{2016}\natexlab{}.
\newblock \showarticletitle{Hardware Trojans: Lessons Learned After One Decade
  of Research}.
\newblock \bibinfo{journal}{\emph{TODAES}} \bibinfo{volume}{22},
  \bibinfo{number}{1} (\bibinfo{year}{2016}), \bibinfo{pages}{6:1--6:23}.
\newblock


\bibitem[Yang et~al\mbox{.}(2016)]%
        {yang16_a2}
\bibfield{author}{\bibinfo{person}{K. Yang}, \bibinfo{person}{M. Hicks},
  \bibinfo{person}{Q. Dong}, \bibinfo{person}{T. Austin}, {and}
  \bibinfo{person}{D. Sylvester}.} \bibinfo{year}{2016}\natexlab{}.
\newblock \showarticletitle{{A2}: Analog Malicious Hardware}. In
  \bibinfo{booktitle}{\emph{Proc. SP}}. \bibinfo{pages}{18--37}.
\newblock


\bibitem[Yasin et~al\mbox{.}(2017)]%
        {yasin17_Anti-SAT}
\bibfield{author}{\bibinfo{person}{M. Yasin} {et~al\mbox{.}}}
  \bibinfo{year}{2017}\natexlab{}.
\newblock \showarticletitle{Security Analysis of {Anti-SAT}}. In
  \bibinfo{booktitle}{\emph{Proc. ASP-DAC}}. \bibinfo{pages}{342--347}.
\newblock


\end{thebibliography}
